\documentclass[fleqn,12pt,twoside]{article}
\usepackage[headings]{espcrc1}
\usepackage{graphicx}
\usepackage{subfigure}
\usepackage[figuresright]{rotating}

\newcommand{\AmS}{{\protect\the\textfont2
  A\kern-.1667em\lower.5ex\hbox{M}\kern-.125emS}}

\title{High-$p_{T}$ ${\pi}^{0},{\eta}$, Identified and Inclusive 
	Charged Hadron Spectra from PHENIX}
\author{
  Maya SHIMOMURA \address{Graduate School of Pure and Applied Sciences, Univ. of Tsukuba, 
  Tenno-dai 1-1-1, Tsukuba, Ibaraki, Japan} (for the PHENIX 
  \thanks{ For the full list of PHENIX authors and acknowledgements, 
    see Appendix 'Collaborations' of this volume.}
      collaboration)
}
\runtitle{High-$p_{T}$ ${\pi}^{0},{\eta}$, Identified and Inclusive 
	Charged Hadron Spectra from PHENIX}
\runauthor{M. SHIMOMURA, for the PHENIX collaboration}

\begin{document}
\maketitle
\begin{abstract}
PHENIX has extended the measurement of the $\pi^{0}$, $\eta$, identified and inclusive charged hadron up to 20 GeV/c, and extended the measurement to the Cu+Cu collision system. 
A strong suppression is observed for both $\pi^{0}$ and charged hadron yields in central Au+Au and Cu+Cu collisions. 
Comparing Au+Au and Cu+Cu systems, $R_{AA}$ becomes independent of $p_{T}$ above 5 GeV/c. Its centrality dependence is compared with two models in order to test for universal $N_{part}$ scaling that is independent of system; results are inconclusive. The results are compatible with energy loss predictions. In addition, the ratio of $\eta$ to  $\pi^{0}$ approaches, within uncertainties, a constant value of  $0.4 \sim  0.5$ at high $p_{T}$ in p+p, d+Au, and Au+Au, while the ratio of $K_{s}$ to $\pi^{0}$ is also consistent with a constant value at high $p_{T}$ in d+Au and p+p. These results are compatible with normal jet fragmentation. 
\end{abstract}

\section{Physics Motivation}
We have previously observed that $\pi^{0}$, $\eta$ and charged hadron yields are significantly suppressed especially for the high $p_{T}$ region ( $p_{T}$ $\geq 4$ $\sim 5$ GeV/c) in Au+Au collision at 200 GeV compared with p+p collisions.\cite{pi0_pp}\cite{pi0_run2}\cite{pi0_h_run2}\cite{h_run2} Since there is no suppression in d+Au collisions at high $p_{T}$,\cite{pi0_h_run2} it is understood that the suppression occurs due to the final state interaction at the collision such as the gluon radiation in the hot dense matter.
Another evidence for the suppression being a final state effect comes from the non-suppression of the direct photon yield in Au+Au collisions.\cite{d_photon}
To understand the character of the suppression more, the comparison between different system size (p+p/d+Au/Cu+Cu/Au+Au) measurements has been studied, and an extended $p_{T}$ reach to 20 GeV/c has been afforded by the long Run4 Au+Au dataset.  
 
\section{Data Analysis, $\pi^{0}$, $K_{s}$, charged hadron, and $\eta$}
We newly measured the following spectra.
 \begin{itemize}
   \item $\pi^{0}$ spectra with extended $p_{T}$ range in Au+Au at 200GeV
   \item $\pi^{0}$ and charged hadron spectra in Cu+Cu at 200GeV
   \item $K_{s}$ spectra in d+Au and p+p at 200GeV \cite{K_s}
   \item Newly finalized $\eta$ spectra in p+p, d+Au, and Au+Au \cite{eta}
 \end{itemize}
The PHENIX experiment consists of four spectrometer arms (two central arms and two muon arms) and a set of global detectors. Each central arm covers the pseudorapidity range $\mid$ $\eta$ $\mid$ $\leq$ $0.35 $ and 90 degrees in azimuth. Charged particles are tracked by a drift chamber (DC) and pad chambers (PC) in each central arms. The electromagnetic calorimeters (EMCal) are used to measure $\gamma$ energy deposit and construct the invariant masses of $\pi^{0}$($\rightarrow 2\gamma$), $K_{s}$($\rightarrow 2\pi^{0} \rightarrow 4\gamma$) and $\eta$($\rightarrow 2\gamma$). \cite{h_run2}\cite{detector}
   
\section{The Nuclear Modification Factor $R_{AA}$}
\subsection{$\pi^{0}$, Charged Hadron and $\eta$}
\begin{figure}[htbp]
\begin{minipage}[h]{0.49\linewidth}
\includegraphics*[width=\linewidth]{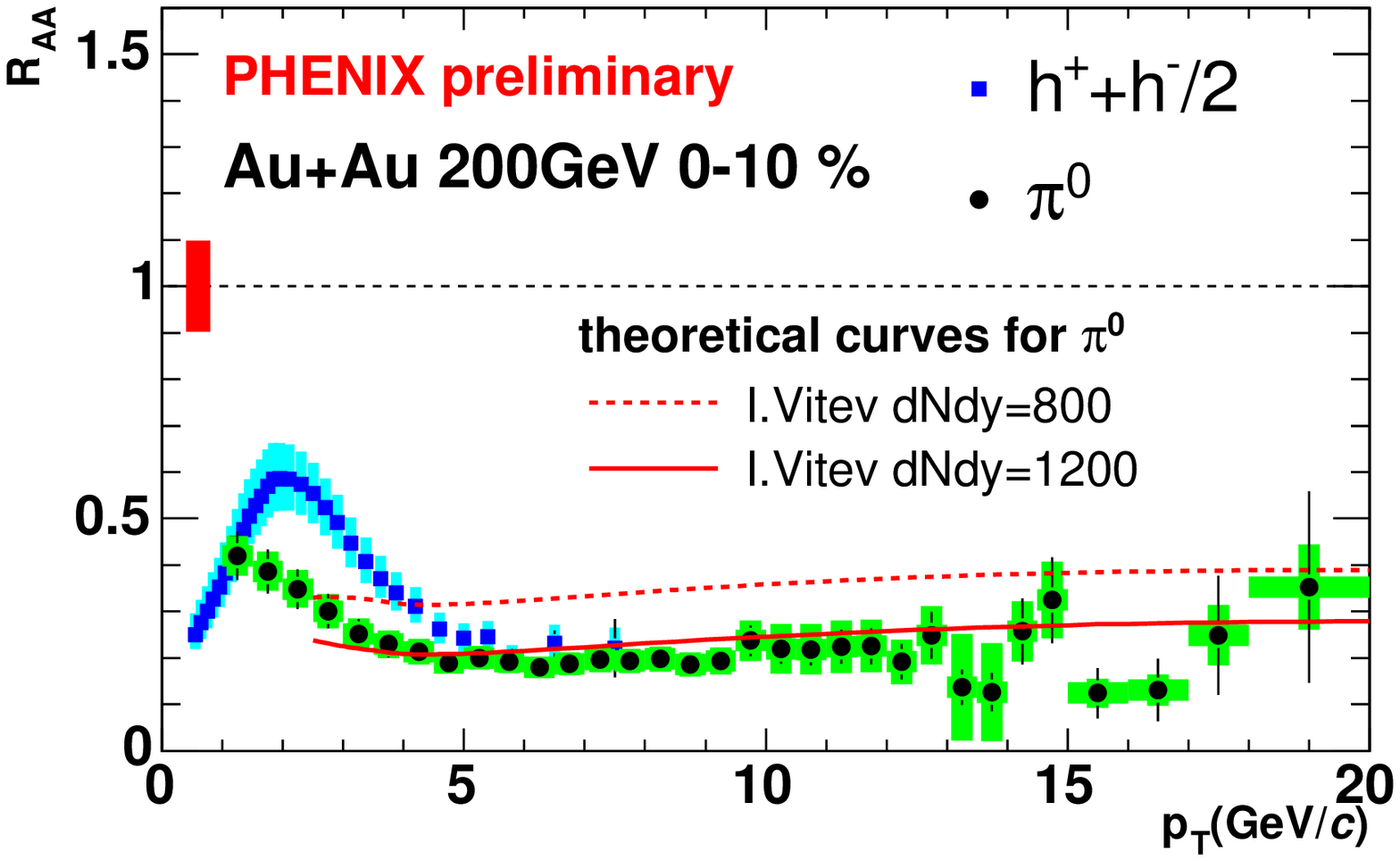}
\end{minipage}
\begin{minipage}[h]{0.49\linewidth}
\includegraphics*[width=\linewidth]{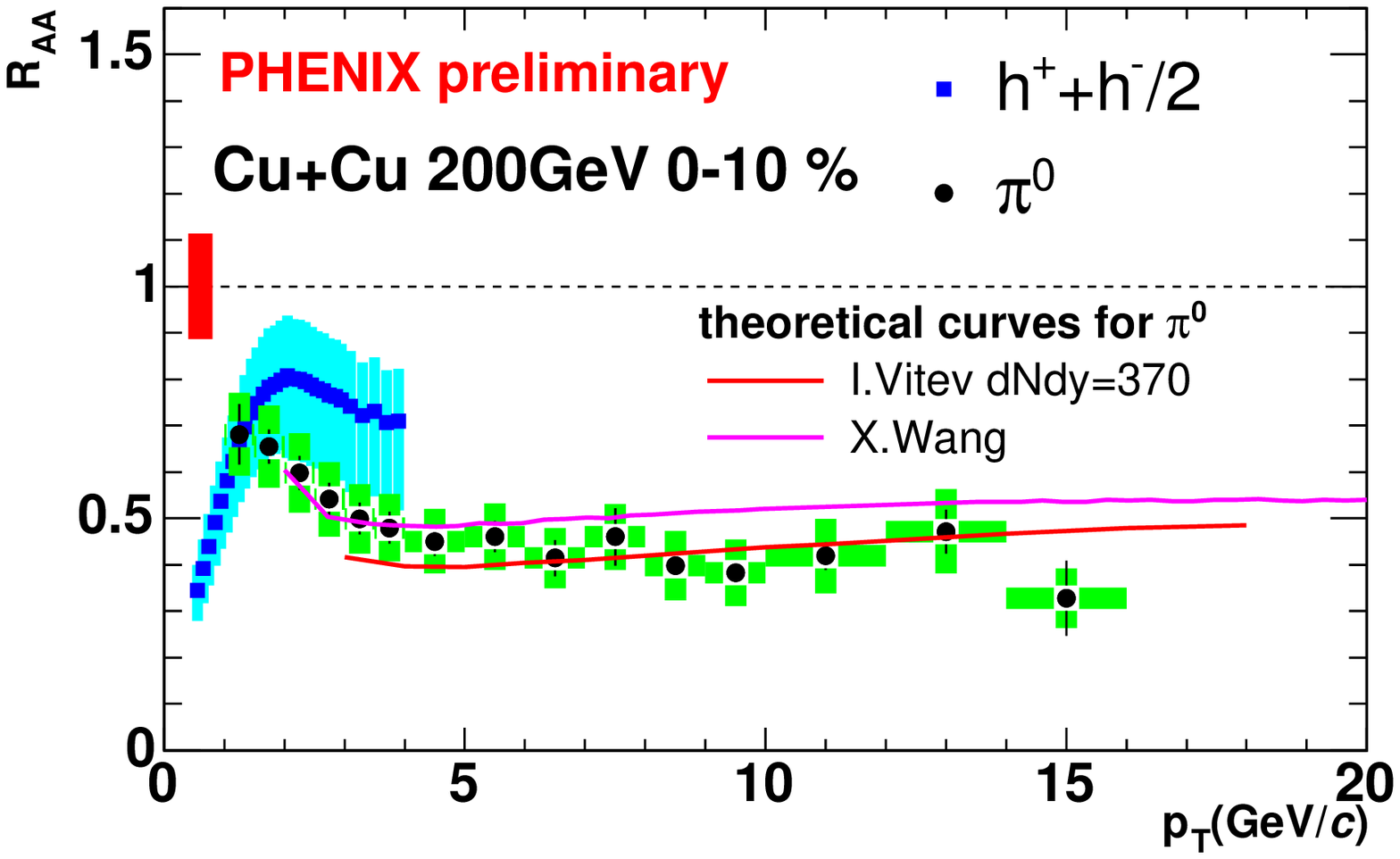}
\end{minipage}
\vskip -10mm
\caption{
      The comparison of $R_{AA}$ for $\pi^{0}$ and 
	charged hadron at 0-10 $\%$ centrality bin in Au+Au (left) and Cu+Cu (right) 
	collisions as a function of $p_{T}$ with theoretical prediction (Red \cite{th_vitev} and purple \cite{energy_loss} lines). 
	The error bars are statistical error, and the boxes are systematic error.
}
\label{fig:Raa}
\end{figure}

The Fig.\ref{fig:Raa}, shows the comparison of $R_{AA}$ for $\pi^{0}$ and charged hadrons in 0-10 $\%$ most central Au+Au and Cu+Cu collisions as function of $p_{T}$. Both $\pi^{0}$ and charged hadron are strongly suppressed in both Au+Au and Cu+Cu collision. The difference between $\pi^{0}$ and charged hadron for $p_{T} \leq 5$ GeV/c comes from the proton contribution. For more central collision, the suppression is getting stronger and the difference between $\pi^{0}$ and charged hadron is getting larger. Above $5$ GeV/c, $\pi^{0} R_{AA}$ becomes flat out to 20 GeV/c. These results are consistent with the model predicting parton energy loss in the medium\cite{energy_loss} and the model predicting shadowing, Cronin effect, and parton energy loss in the medium.\cite{th_vitev} In addition, $\eta$ is also suppressed in central Au+Au collisions and the suppression pattern is similar to $\pi^{0}$. \cite{eta}   
\vskip -6mm

\vskip -20mm
\subsection{Comparison between Au+Au and Cu+Cu}
\begin{figure}[htbp]
\subfigure[]{

\includegraphics*[width=0.45\linewidth]{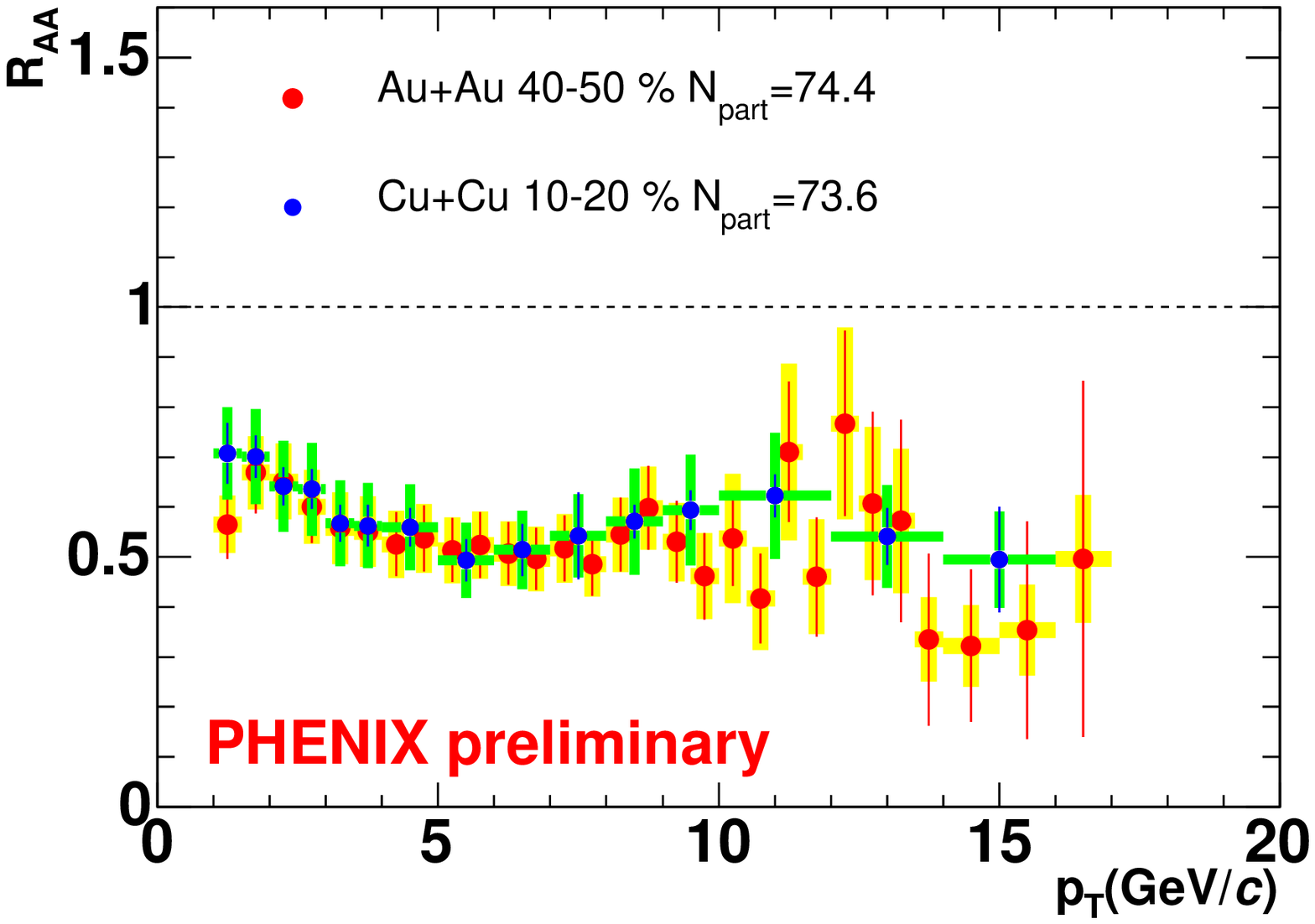}
}
\hspace*{0.1\linewidth}
\subfigure[]{
\includegraphics*[width=0.35\linewidth]{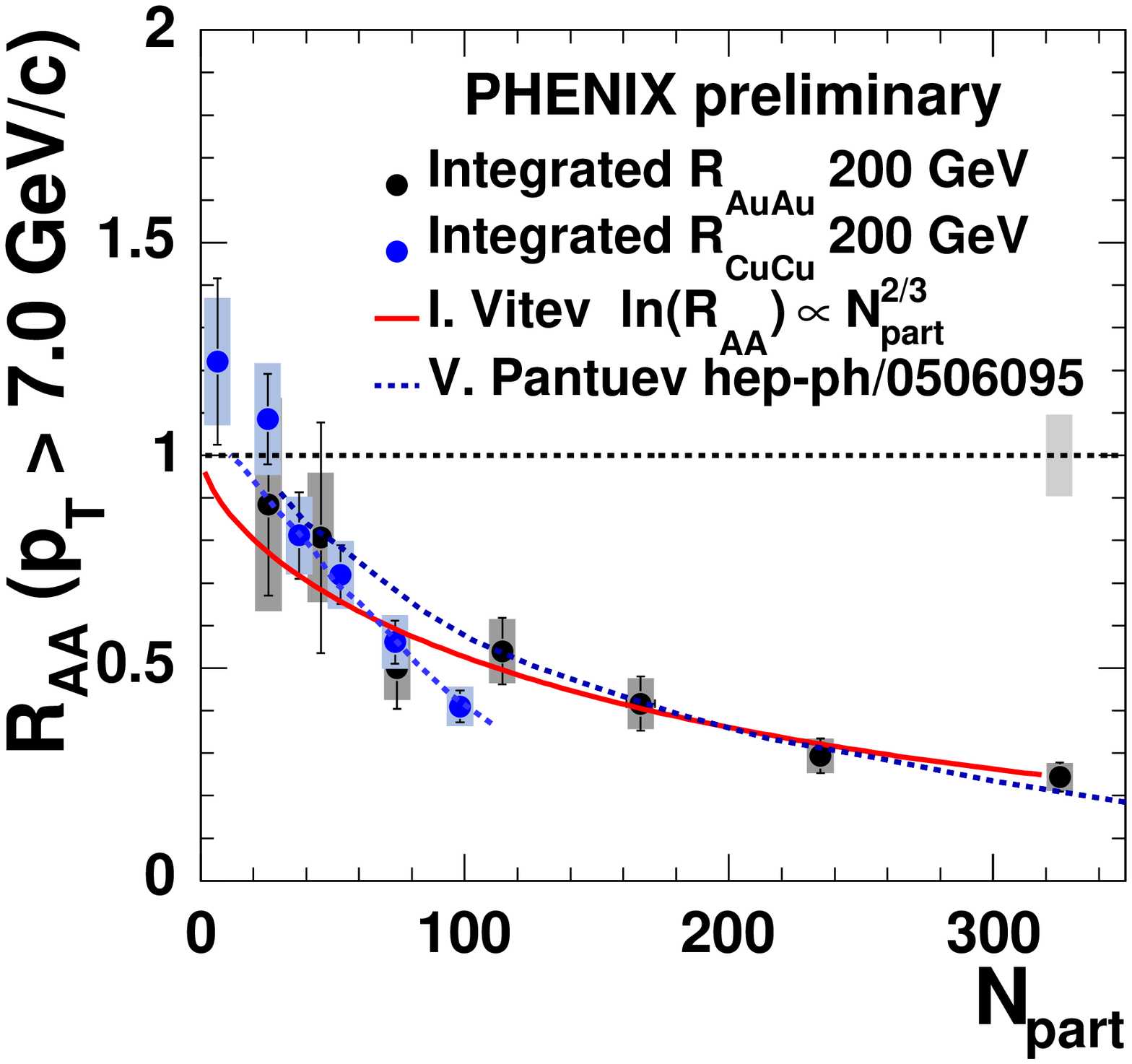}
}
\vskip -10mm
\caption{(a)The comparison between $\pi^{0}$ $R_{AA}$ in Au+Au and Cu+Cu at similar $N_{part}$ ($\sim 74$). (b)The integrated $R_{AA}$ at $7 \leq p_{T} \leq 20$ GeV/c with theoretical curves \cite{th_patuev}\cite{th_vitev} as a function of $N_{part}$.}
\label{fig:Raa_theory}
\end{figure}

In Fig.\ref{fig:Raa_theory}(a), the comparison of $R_{AA}$ in Au+Au to that in Cu+Cu with similar $N_{part}$ is shown. The suppression is similar for similar $N_{part}$ at mid-centrality. Though a universal $N_{part}$ scaling is roughly suggested,\cite{th_jjia} due to the size of the uncertainties, it is unclear whether a single scaling curve can exactly describe the suppression ($R_{AA}$) in both systems simultaneously. In Fig.\ref{fig:Raa_theory}(b), we show the integrated $R_{AA}$ of $\pi^{0}$ at  $7.0 \leq p_{T} \leq 20.0$ GeV/c with two different theoretical curves \cite{th_patuev}\cite{th_vitev} as a function of $N_{part}$. Both models are consistent with the data from central to mid-central collisions.
 
\vskip -15mm
\section{The Particle Ratio}
\begin{figure}[htbp]
\begin{minipage}[h]{0.49\linewidth}
\includegraphics*[width=\linewidth]{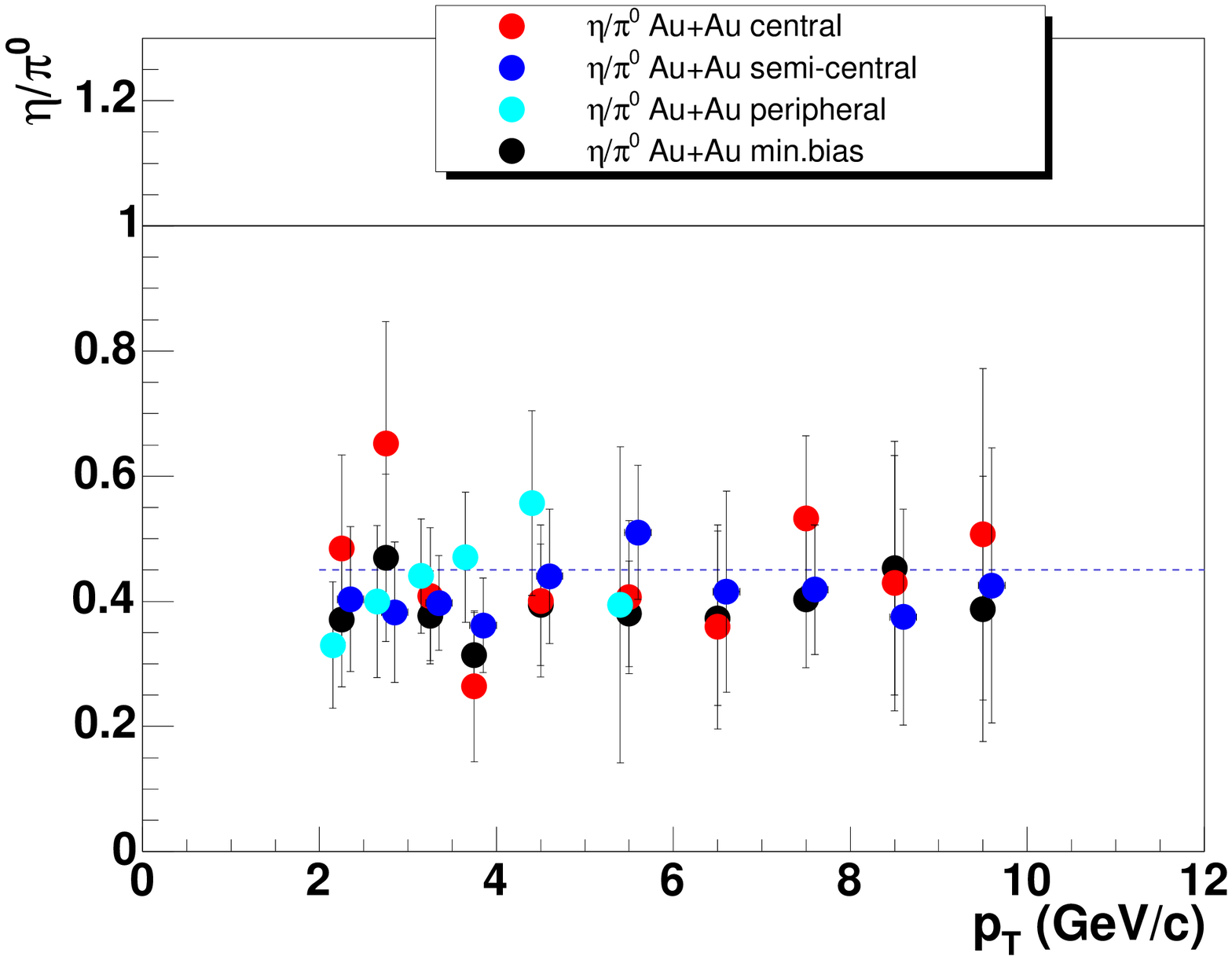}
\end{minipage}
\begin{minipage}[h]{0.49\linewidth}
\includegraphics*[width=\linewidth]{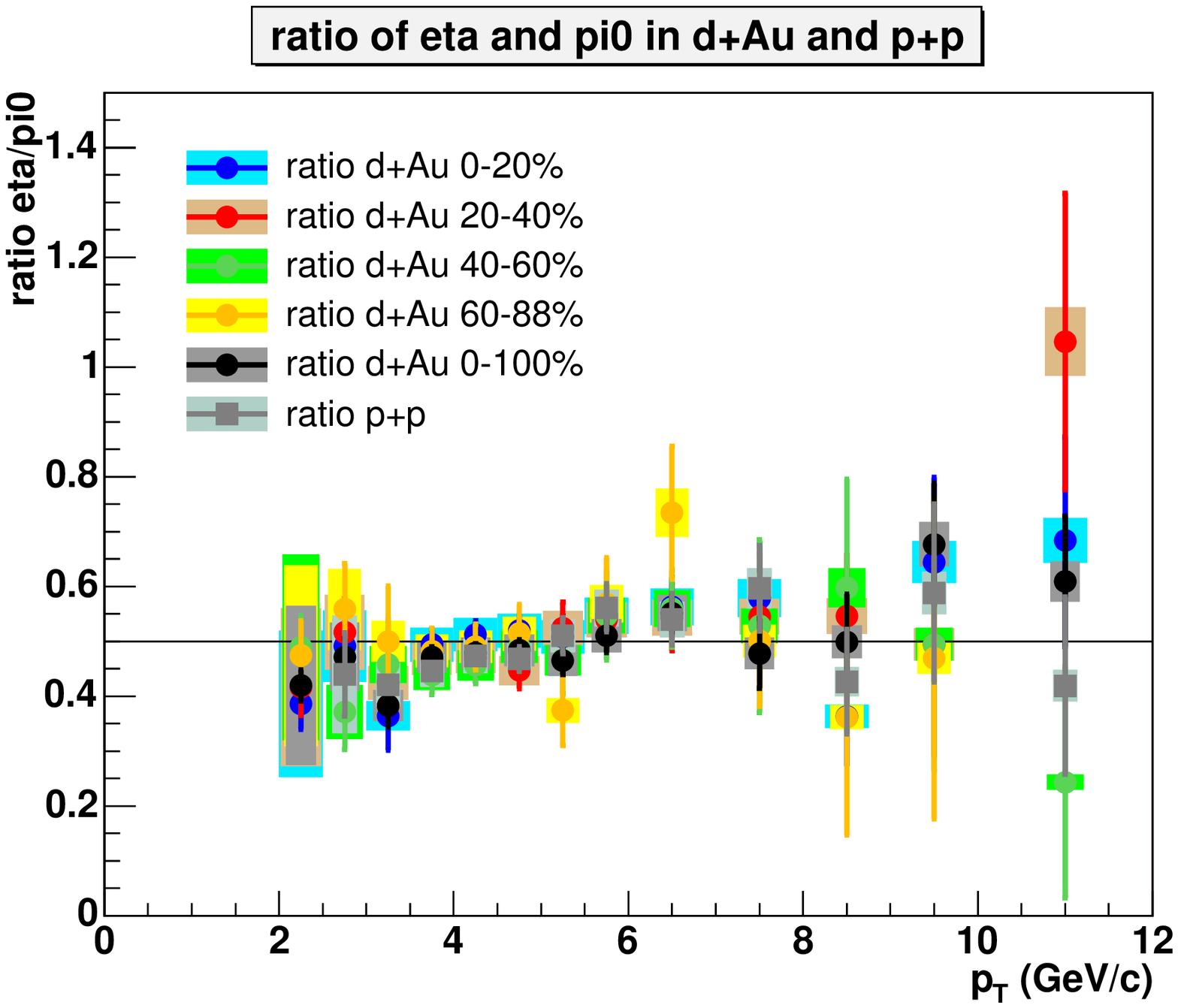}
\end{minipage}
\vskip -8mm
\caption{
	The ratio of $\eta$ to $\pi^{0}$ in Au+Au (left), d+Au (right) and p+p (right) as a function of $p_T$ at $\sqrt{s} = 200$ GeV. 
	The error bars are statistical error, and the boxes are systematic error.
}
\label{fig:ratio_eta}
\end{figure}

\begin{figure}[htbp]
\begin{minipage}[h]{0.40\linewidth}
\includegraphics*[width=\linewidth]{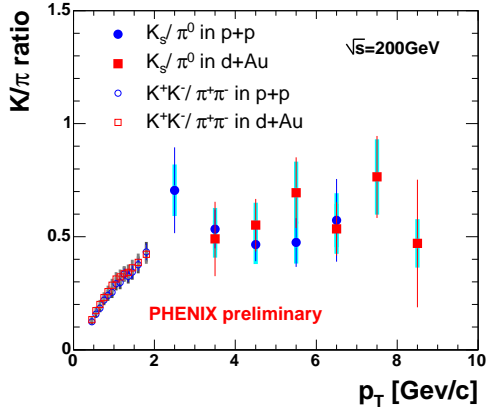}
\end{minipage}
\hspace*{0.05\linewidth}
\begin{minipage}[h]{0.45\linewidth}
\caption{
      The closed symbols are the ratio of $K_{s}$ to $\pi^{0}$ in p+p and d+Au as a function of $p_T$. The opened symbols show the ratio of of $K^{\pm}$ to $\pi^{\pm}$ as reference.\cite{K_pi_ratio} 
	The error bars are statistical error, and the boxes are systematic error.
}
\label{fig:ratio_ks}
\end{minipage}
\vskip -10mm
\end{figure}

The ratio of $\eta$ to $\pi^{0}$ is $\sim 0.4 - 0.5$ in all systems and for all centralities as shown in fig.\ref{fig:ratio_eta}.\cite{eta}. Also, the ratio of $K_{s}$ to $\pi^{0}$ at p+p and d+Au becomes flat at high $p_{T}$ as shown in fig.\ref{fig:ratio_ks}.\cite{K_s}  Therefore, the mesons are affected by the medium in the same way in different collision systems. These results are consistent with jet fragmentation at high $p_{T}$.  
\vskip -5mm

\vskip -10mm
\section{Summary}
 We have studied $\pi^{0}$,$\eta$, $K_{s}$ and charged hadron spectra in Au+Au, Cu+Cu, d+Au and p+p at high $p_{T}$. For $\pi^{0}$ and charged hadron, we observed the suppressions in both Cu+Cu and Au+Au collisions compared with p+p collisions, and no suppression is observed in d+Au collisions. The $R_{AA}$ comparison between Au+Au and Cu+Cu indicates that the suppression is almost the same for similar $N_{part}$.  A universal $N_{part}$ scaling of $R_{AA}$, independent of system, describes the data in an approximate sense, but it cannot be confirmed exactly due to experimental uncertainties. In addition, the high $p_{T}$ $\pi^{0}$ suppression is flat out to 20 GeV/c and its magnitude is quantitatively consistent with parton energy loss model calculations. $\eta$ has a similar suppression pattern as $\pi^{0}$ does. The ratio of $\pi^{0}$ to $\eta$ is independent of centralities and system size. Similarly, the $K_{s}$ to $\pi^{0}$ ratio is constant within uncertainties at high $p_{T}$ for both p+p and d+Au. These particle ratios are consistent with normal jet fragmentation.

\end{document}